\documentclass[preprint,12pt,3p]{elsarticle}

\usepackage{graphicx}
\usepackage{epstopdf,epsfig}
\usepackage{amssymb}
\usepackage{amsmath}
\usepackage{makecell}
\usepackage{tikz}
\usepackage{dirtytalk}
\usepackage{float}
\usepackage{tabularx}
\usepackage{adjustbox}
\usetikzlibrary{shapes}

\DeclareRobustCommand\sampleline[1]{%
\tikz\draw[#1] (0,0) (0,\the\dimexpr\fontdimen22\textfont2\relax)
-- (1.6em,\the\dimexpr\fontdimen22\textfont2\relax);}

\let\today\relax

\begin{document}
\begin{frontmatter}

\title{ \vspace{-0.25in} {\footnotesize \em ICLASS 2018, 14th Triennial International Conference on Liquid Atomization and Spray Systems, Chicago, IL, USA, July 22-26, 2018} \\ Volumetric Displacement Effects of Dispersed Phase on the Euler-Lagrange Prediction of a Dense Spray}

\author[label1]{Pedram Pakseresht}
\address[label1]{School of Mechanical, Industrial and Manufacturing Engineering, Oregon State University, Corvallis, OR 97331, USA}\ead{pakserep@oregonstate.edu}

\author[label1]{Sourabh V. Apte\corref{cor1}}
\ead{Sourabh.Apte@oregonstate.edu}

\cortext[cor1]{Corresponding author. 204 Rogers Hall, Corvallis, OR 97331, USA. Tel: +1 541 737 7335, Fax: +1 541 737 2600.}

\begin{abstract}
Accurate prediction of a dense spray using an Euler-Lagrange approach is challenging because of high volume fraction of the dispersed phase due to subgrid cluster of droplets. To accurately model dense sprays, one needs to capture this effect by taking into account the spatio-temporal changes in the volume fraction of the carrier phase due to the motion and presence of the dispersed phase. This leads to zero-Mach number, variable density equations which are commonly neglected in the standard two-way coupling spray simulations. Using pressure-based solvers, this gives rise to a source term in the pressure Poisson equation and a non-divergence free velocity field. To validate the predictive capability of such approach, an atomized non-evaporating dilute particulate round jet is first examined using Large Eddy Simulation coupled with Point-Particle approach and then higher volume loadings up to $38\%$ are investigated with and without taking into account the volumetric displacement effects. It is shown that for volume loadings above $5\%$, the volumetric displacement effects enhance dynamics of the flow resulting in a higher stream-wise mean and r.m.s. velocities compared to the results of standard two-way coupling. This is more pronounced for the near field of the jet where local volume fraction of the dispersed phase is relatively high. This enhancement is conjectured to be due to the velocity divergence effect due to the modified continuity equation where spatio-temporal variations in volume fraction of the carrier phase increases velocity in the regions of high void fraction.  
\end{abstract}

\end{frontmatter}

\section{Introduction}
Liquid spray atomization plays an important role in analyzing the combustion process in many propulsion related applications. Disintegration of injecting liquid fuel which occurs in two steps; primary atomization followed by secondary atomization makes this process a complex flow. This has opened a new field of study for investigating and modeling this flow. In the traditional approaches for spray modeling, the dynamics of the liquid/air interface are not resolved. Instead, the liquid phase is modeled through either an Eulerian approach in which droplets are considered to be as a continuous liquid phase or Lagrangian Point-Particle/Parcel (PP) method where droplets are assumed subgrid and their motion is captured by force closures such as drag, buoyancy, pressure, etc., while the effect of the droplets on the gas phase is modeled through two-way coupling of mass, momentum, and energy exchange \cite{Dukowicz1980}. In such scenario, volume fraction of droplets compared to the computational cell must be very small so that the Point-Particle approach could be practical and feasible. This restriction, however, prevents Euler-Lagrange approaches from applying to regions with high void fractions such as near the nozzle exit where primary atomization takes place. This led researchers to perform Eulerian-Lagrangian Spray Atomization (ELSA) approach which couples the Eulerian mixing description for primary atomization (LES/DNS) with Lagrangian formulation (PP method) for secondary breakup \cite{Blokkeel2003,Lebas2005}. These models were originally derived in the context of RANS turbulence models and assume infinite Weber number, however, extensions to LES formulations have been recently proposed by \cite{Chesnel2011a,Chesnel2011b}. Recently, hybrid approaches of DNS method for the primary atomization region along with LES coupled with Lagrangian Point-Particle/Parcel approach for solving the gas and liquid phases respectively in the secondary atomization region have been developed by, e.g., \cite{Herrmann2010a,Herrmann2011} among others. These hybrid approaches have shown quite success in predicting atomization even in complex aircraft engine injectors, yet on the one hand they are still computationally expensive. On the other hand, for the secondary atomization, the spatio-temporal variations in volume fraction of the carrier phase is still neglected in Euler-Lagrange approaches. This effect can be significant particularly in atomizing sprays where volume fraction of the injecting liquid phase is in the order of one near the nozzle exit. Therefore, ignoring this effect would eliminate the real dynamics of such flows in which the liquid phase displaces a remarkable portion of the gaseous phase. 

Therefore, in this work, to take advantage of computationally less expensive Euler-Lagrange approaches as well as for more accurate predictions of dense spray flows, an Euler-Lagrange approach modified with spatio-temporal variations in the carrier phase volume fraction is applied \cite{Anderson1967,Dukowicz1980,Joseph1990}. In order to distinguish this method from the standard Euler-Lagrange two-way coupling approaches, we refer to this model as volumetric coupling in line with work of \cite{Cihonski2013}. Although there have been some studies showing the insignificance of volumetric coupling compared to the standard two-way coupling for dilute regimes (e.g., \cite{Vreman2004}), several works have depicted the importance of this unique coupling. \cite{Apte2008} illustrated a large difference in prediction of particle dispersion in fluidization process using this approach compared to the standard Euler-Lagrange approach. \cite{Ferrante2004} observed that taking into account the volumetric displacement effect results in accurately capturing drag reduction in the boundary layer over a flat plate. In line with previous works, \cite{Cihonski2013} showed that even under dilute loading (small number of bubbles entrained in a vortex ring), accounting for this effect can significantly alter the vortex core for certain combinations of the vortex strengths and bubble sizes. In the current work, similar formulation is applied to an atomized non-evaporating particulate jet flow to only focus on the volumetric displacement effect rather than atomization processes. We start looking a dilute particulate round jet as a validation case and then increasing the corresponding volume loading of the dispersed phase (here solid particles) up to 38\% while keeping other flow parameters constant. Four different volume loadings are studied here to investigate the importance of volumetric displacement effect at different loadings explained in the next sections.

\section{Methodology}
\label{Methodology}
An Euler--Lagrange approach is used to simulate the particle-turbulence interactions involved in dense spray flows. Fluid motion is captured through solving the Navier-Stokes as well as continuity equations in an Eulerian framework using large eddy simulation (LES) while motion of particles is modeled in a Lagrangian framework using available force closures, the so-called Point-Particle approach \cite{Maxey1987, Elghobashi1991, Squires1991}. Governing equations on these two phases are explained in the following sections in detail.

\subsection{Dispersed Phase Formulation}
Small particles (i.e., smaller than the resolved fluid length scale) are tracked through the Newton second law of motion given the forces exerted by fluid as shown in Equation \ref{newton} \cite{Maxey1983}. Unlike body-fitted approaches (e.g. \cite{pakseresht2012}), the no-slip condition on the surface of particles is not imposed in this approach. According to this, one can obtain velocity components, $\mathbf{u}_p$, as well as position, $\mathbf{x}_p$, of each individual particle of mass $m_p$. 

\begin{equation}
\begin{split}
\frac{d}{dt}(\mathbf{x}_p) &= \mathit{\mathbf{u}_p} \\ 
\frac{d}{dt}(\mathbf{u}_p) &= \frac{1}{m}_p (\mathbf{F}_g  + \mathbf{F}_{pr} + \mathbf{F}_d +  
\mathbf{F}_{l,Saff} + \mathbf{F}_{l,Mag} + \mathbf{F}_{am})  
\label{newton} 
\end{split}
\end{equation}

Equation \ref{newton} shows all possible forces including gravitational body force, $\mathbf{F}_g$, hydrostatic pressure gradient, $\mathbf{F}_{pr}$, shear induced lift force \cite{Saffman1965}, $\mathbf{F}_{l,Saff}$, Magnus effect due to particle rotation \cite{Rubinow1961}, $\mathbf{F}_{Mag}$, as well as added mass \cite{Auton1983}, $\mathbf{F}_{am}$. In order to capture more accurately particle-turbulence interactions in a dense spray regime, the drag closure by \cite{Tenneti2011}, $\mathbf{F}_d$, is employed here where coefficient of drag, $C_d(Re_p,\theta_p)$ accounts for local volume fraction of dispersed phase as well as finite particle Reynolds number. It has been observed that the Basset history force does not remarkably affect motion of particles in the presence of steady drag force \cite{Maxey1983,Bagchi2003}, therefore, this force is excluded in this study. All aforementioned forces above are given in Equations \ref{gravity}-\ref{drag} as follow; 
 
\begin{equation}
\mathbf{F}_g = (\rho_p-\rho_g)V_p\mathbf{g}  \quad ; \quad g = -9.81m/s^2
\label{gravity} 
\end{equation}

\begin{equation}
\mathbf{F}_{pr} = -V_p \nabla P_{|p}
\label{pressure} 
\end{equation}

\begin{equation}
\mathbf{F}_{l,Saff} = m_pC_l \frac{\rho_f}{\rho_p}(\mathbf{u}_{f|p} -\mathbf{u}_p)\times (\nabla \times \mathbf{u}_f)_{|p}, \quad \text{and} \quad C_l = \frac{1.61\times6}{\pi d_p}\sqrt{\frac{\mu_f}{\rho_f}|(\nabla\times \mathbf{u}_f)_{|p}|}
\label{saffman} 
\end{equation}

\begin{equation}
\mathbf{F}_{l,Mag} = C_{mag}\frac{\mathbf{u}_{rel} \times \mathbf{\Omega}_{rel}}{|\mathbf{\Omega}_{rel}|}(\frac{1}{2}\rho_f |\mathbf{u}_{rel}|A), \quad \text{and} \quad C_{mag} = min(0.5,0.25\frac{d_p|\omega_{rel}|}{|u_{rel}|})
\label{magnus}
\end{equation}

\begin{equation}
\mathbf{F}_{am} = m_pC_{am} \frac{\rho_f}{\rho_p}(\frac{D\mathbf{u}_{f|p}}{Dt} - \frac{d\mathbf{u}_p}{dt}), \quad C_{am} = 0.5
\label{addmass} 
\end{equation}

\begin{equation}
\centering
\begin{split}
&\mathbf{F}_d  = m_p \frac{C_d(Re_p,\theta_p)}{\tau_p}(\mathbf{u}_{f|p} -\mathbf{u}_p) \\
&C_d(Re_p,\theta_p) = (1-\theta_p)(\frac{C_d(Re_p,0)}{(1-\theta_p)^3} + A + B),\\
&A = \frac{5.81\theta_p}{(1-\theta_p)^3}+0.48\frac{\theta^{1/3}}{(1-\theta_p)^4}, \\
&B = \theta^3_p Re_p(0.95+\frac{0.61\theta^3_p}{(1-\theta_p)^2}), \\
&C_d(Re_p,0) = 1 + 0.15 Re^{0.687}_p.
\label{drag}
\end{split}     
\end{equation}

\noindent where volume and volume fraction of each particle are represented by $V_p$ and $\theta_p$ respectively. $\mathbf{u}_{rel}= \mathbf{u}_{f|p}-\mathbf{u}_p$ is the relative velocity between fluid (seen by particle) ($\mathbf{u}_{f|p}$) and particle ($\mathbf{u}_p$). On the other hand, $\tau_p=(\rho_p{d_p}^2)/(18\rho_f \mu_f \theta_f)$ and $Re_p=(\theta_{f|p} \rho_f |u_{rel}|d_p)/(\mu_f)$ are the respective particle relaxation time and particle Reynolds number modified by local volume fraction of the dispersed phase \cite{Finn2016}.

\subsection{Fluid Phase Formulation}
Large eddy simulation (LES) is employed to solve the carrier phase equations in a structured Cartesian grid using finite volume discretization. A pressure based second order fractional time step method based on work of \cite{Finn2011} and \cite{Cihonski2013}, adjusted to a co-located structured grid by \cite{Finn2016} is utilized here. To consider the effect of volume of carrier phase displaced by the motion and presence of particles, the volume filtered Navier-Stokes equations given in (\ref{mass})-(\ref{momentum}) are applied here \cite{Anderson1967, Joseph1990}. Despite typical incompressible flows, the spatio-temporal variations in particle volume fraction generate a non-divergence free velocity field in the flow.

\begin{equation}
\frac{\partial (\rho_f \theta_f)}{\partial t} + \nabla \cdot (\rho_f \theta_f\mathbf{u}_f) = 0.
\label{mass}
\end{equation} 

\begin{equation} 
\frac{\partial (\rho_f \theta_f \mathbf{u_f} ) }{\partial t} + \nabla \cdot ( \rho_f \theta_f \mathbf{u_f} \mathbf{u_f}) =  -\theta_f \nabla P + \nabla \cdot [\mu_f\theta_f(\nabla \mathbf{u}_f + \nabla \mathbf{u}^T_f)]+ \theta_f \rho_f \mathbf{g} + \mathbf{F}_{p \rightarrow f}.
\label{momentum}
\end{equation}

\noindent where $\rho_f$, $\theta_f$, and ${\bf{u}}_f$ are density, volume fraction, and velocity of the carrier phase respectively. Volume fraction of carrier phase is calculated as $\theta_f = 1 - \theta_p$, where $\theta_p$ corresponds to volume fraction of particles located in each computational cell. The point-particle forces, $\mathbf{F}_{p \rightarrow f}$, include the equal and opposite reaction forces from the particle surface forces except the pressure force. Accordingly, it is crucial to define a function to project Lagrangian quantities of dispersed phase back to the continuous field as well as interpolate carrier phase properties to the particles' position. Gaussian function \cite{Apte2008} for both interpolation and projection purposes is employed with a dynamic bandwidth equal to the local grid size of each particle's neighbourhood. It is worth mentioning that given $\theta_f=1$, the above formulation switches to the standard two-way coupling where volumetric displacement effects of the carrier phase is then neglected. For large-eddy simulation, the equations above should be spatially filtered using density weighted Favre averaging \cite{hutter2004}. For turbulent flows, the Favre averaged equations then have the same form as Equation \ref{momentum} with the exception that the left-hand side of the momentum equation consists of an unclosed subgrid stress term, 
 
\begin{equation} 
\tau_{ij} = \overline{\rho_f\theta_fu_iu_j} - \overline{\rho_f\theta_fu_i} \:  \overline{\rho_f\theta_fu_j}/\overline{\rho_f\theta_f}
\label{sgs}
\end{equation} 

\noindent denoting $\rho=\rho_f\theta_f$, Favre-averaged velocity field can be obtained as $\overline{\rho}\tilde{u_i} = \overline{\rho u_i}$. The unclosed subgrid-scale stresses can be closed by use of a Smagorinsky model with a dynamic procedure for the calculation of the model coefficient and eddy viscosity \cite{germano1991,moin1991}.

\begin{equation} 
\mu_T = -C_S\overline{\rho_f\theta_f}\Delta^2S(\tilde{u}) \quad ; \quad \Delta = V_{cv}^{1/3} \quad ; S(\tilde{u}) = (\frac{1}{2}S_{ij}S_{ij})^{1/2}
\end{equation} 

\noindent where $V_{cv}$ is the volume of a grid cell, and the model constant $C_S$ is obtained using the dynamic procedure and a test filter of twice the size of the control volume. The governing equation then is obtained in terms of the filtered velocity fields $\tilde{u}$ and the fluid viscosity $\mu_f$ us changed to $\mu_{eff}=\mu_f+\mu_T$ accounting for the eddy viscosity. 

\section{Numerical Results} 
 \label{sec:results}
The numerical scheme explained above has been widely applied to and validated for different applications \citep{Shams2011,Cihonski2013,Pakseresht2014,Pakseresht2015,Pakseresht2016,Finn2016,Pakseresht2017_ASME,Pakseresht_2017_aps,pakseresht2018,he2018effect}. Besides, in this work the accuracy and robustness of the method is also enforced on an atomized non-evaporating turbulent round jet laden with dilute regime of inertial particles. Then, to investigate the volumetric displacement effects of dispersed phase, four denser cases, i.e., B-E shown in Table \ref{tab:cases} are performed. In these cases all parameters except volume loading are the same as those in case A. Case E is similar to case D yet instead of two-way coupling, the momentum force exchange term is excluded. As listed in Table \ref{tab:cases}, for cases B-D, the volumetric coupling is compared to the corresponding standard two-way coupling while for case E one-way coupling is compared to its modification by volumetric displacement effects (volumetric formulation).  An experimental data \cite{Mostafa1989} is only available for case A while no data is available for higher loadings. This case validates our numerical results. For further results on the validation see \cite{pakseresht_ilass17_arxiv}. For other cases, only numerical results from different couplings are compared together to investigate the volumetric displacement effects. In all studied cases here, a Cartesian structured grid is used for solving the flow in a rectangular computational domain with size of $6d_{jet}\times6d_{jet}\times14d_{jet}$ in cross-sectional and longitudinal directions, respectively. Inflow data over several flow through times is generated a priori and read at each flow time step to specify the fluid velocity components at the inlet. Convective outflow boundary condition is applied at the oultlet while slip boundary condition is enforced for other sides of the computational domain. As given in table \ref{tab:cases}, particles of 105 micron with specific gravity, $S.G.=\rho_p/\rho_f$, of 2122.24 are injected at the nozzle exit based on different volume loadings. Reynolds number of clear jet denoted by $Re_j=\rho_fU_jd_j/\mu_f=5712$ is defined based on the jet bulk velocity, $U_j=3.546 m/s$, jet nozzle diameter, $d_j=0.0253 m$, and the carrier phase properties for air, $\rho_f=1.178 kg/m^3$ and $\mu_f=1.8502\times10^{-5} kg/(m.s)$ corresponding to the experimental work of \cite{Mostafa1989}.  
 
\begin{table}
\centering
\begin{adjustbox}{width=\textwidth}
\begin{tabular} {lcccccccc}
\hline
Case & $d_p(\mu m)$ & $Re_{j}$ & S.G. & $[\overline{\theta_p}]_{inlet}$ & $N_p (million)$ & Inter-particle collision & Coupling type\\ \hline 
A & 105 &  5712 & 2122.24 & 0.00047 & 0.15 & No & 2-way w/wo volumetric effect\\ 
B & 105 &  5712 & 2122.24 & 0.047 & 0.34 & Yes & 2-way w/wo volumetric effect\\ 
C & 105 &  5712 & 2122.24 & 0.188 & 1.3 & Yes & 2-way w/wo volumetric effect\\ 
D & 105 &  5712 & 2122.24 & 0.376 & 2.56 & Yes & 2-way w/wo volumetric effect\\ 
E & 105 &  5712 & 2122.24 & 0.188 & 1.3 & Yes & 1-way w/wo volumetric effect \\ \hline
\end{tabular}
\end{adjustbox}
\caption{Particle and flow properties for different cases.}
\label{tab:cases}
\end{table}

\begin{figure}
\begin{minipage}{75mm}
\begin{center}
\includegraphics[width=7cm,keepaspectratio=true]{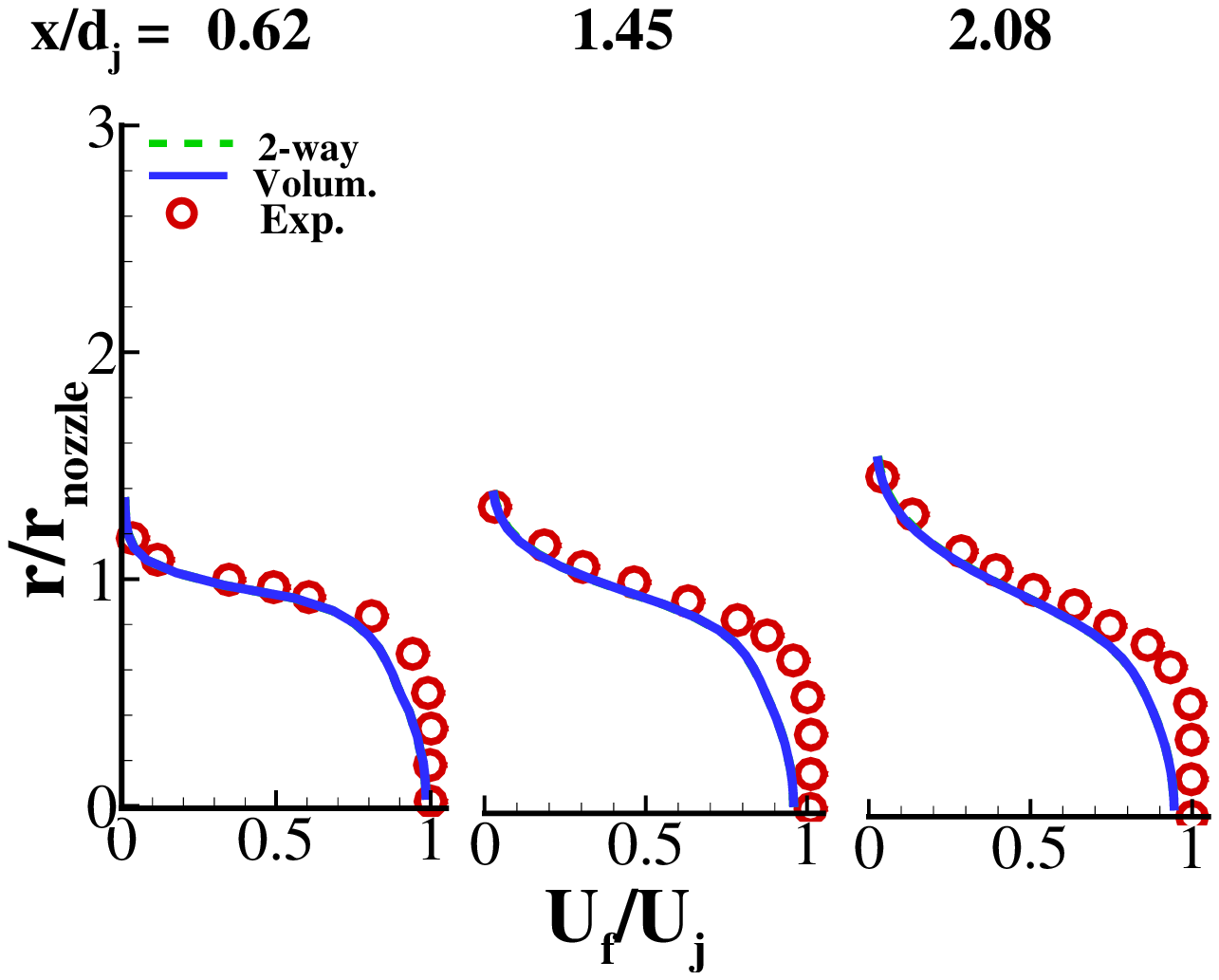}
\caption{Normalized mean velocity of the carrier phase based on two different couplings compared with experimental data of \cite{Mostafa1989}}
\label{fig:mean_f1_only3}
\end{center}
\end{minipage}
\hfill
\begin{minipage}{75mm}
\begin{center}
\includegraphics[width=7cm,keepaspectratio=true]{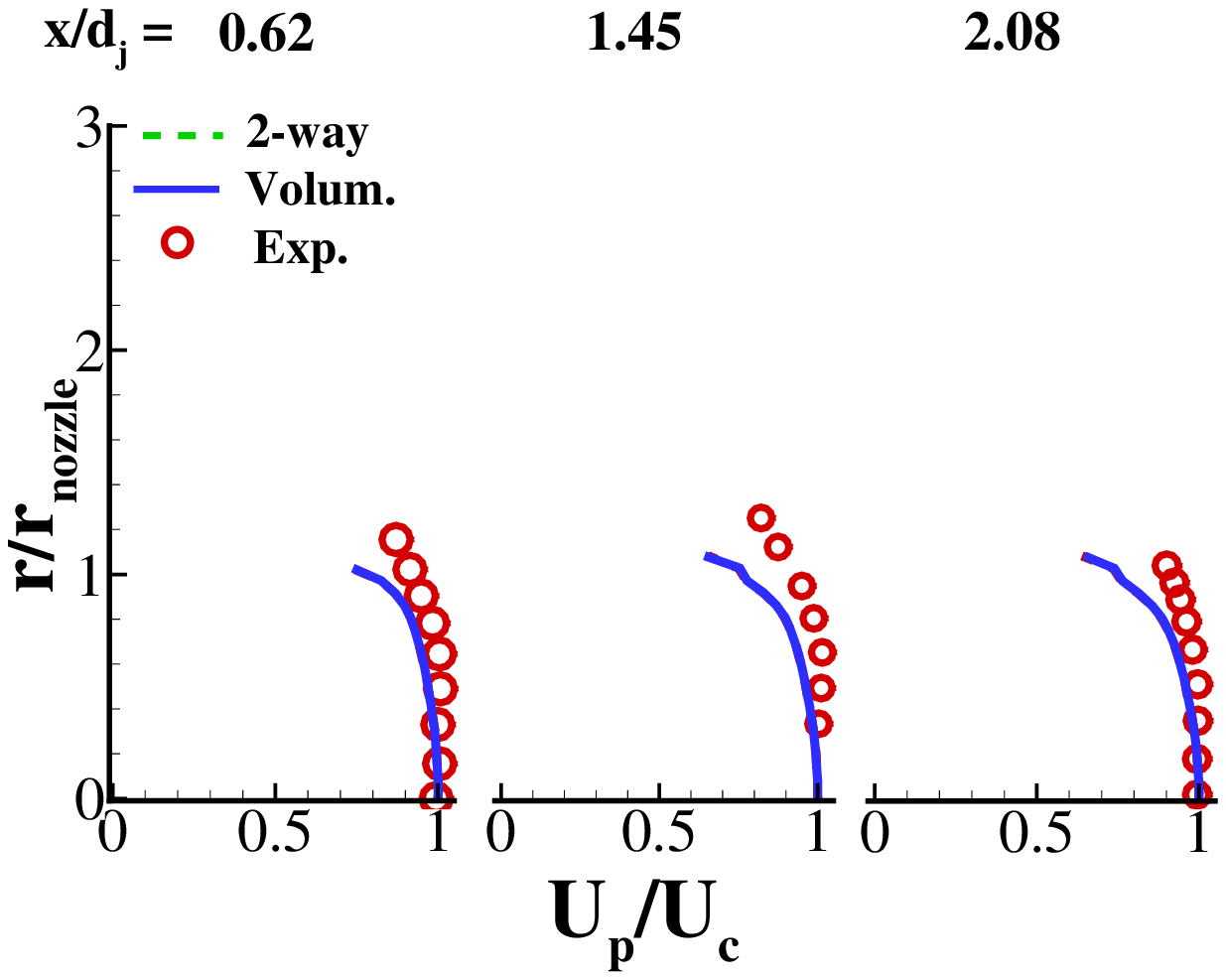}
\caption{Normalized mean velocity of the dispersed phase based on two different couplings compared with experimental data of \cite{Mostafa1989}}
\label{fig:mean_p1_only3}
\end{center}
\end{minipage}
\end{figure}

Regarding dispersed phase, measured data at $x/d_{jet}=0.04$ in the corresponding experiment of case A is used to prescribe mean and r.m.s. velocities in all simulations for injecting particles in the already statistically stable turbulent jet flow. Numerical results at each nozzle distance are obtained based on the azimuthal averaging in space as well as long enough time averaging. Total number of injecting particles for each case required at each time step ($\Delta t_f$) can be calculated based on the given volume loading, particle's diameter and jet bulk velocity as $n_p = 6[\overline{\theta_p}]_{inlet}{d_j}^2U_j \Delta t_f/4 {d_p}^3$. Figures \ref{fig:mean_f1_only3} and \ref{fig:mean_p1_only3} show the comparison of our numerical simulation with experimental data for both carrier and dispersed phases of case A respectively. All results are plotted in normalized radial direction based on jet radius of $r_j$. Mean velocities of the carrier and dispersed phases are normalized with bulk velocity of the clear jet ,$U_j$, and local centreline velocity of the laden jet at each nozzle distance, $U_c$, respectively. A good agreement between numerical results and corresponding experiment on mean velocity of two phases is achieved. On each plot, volumetric coupling prediction matches the results of standard two-way coupling. That would be expected as loading of the dispersed phase in case A is so dilute that even inter-particle collision is negligible and thus the dominant mechanism is only the two-way coupling \cite{Elgobashi2006}. This was also observed by \cite{Vreman2004} where they found insignificant difference between two-way and volumetric couplings for a channel flow laden with average solid volume fraction of $\overline{\theta_p} = 0.013$ (i.e., 1.3\%).  

\subsection{Volumetric Displacement Effects}
\begin{figure}
\setlength{\unitlength}{0.012500in}
\centering
\includegraphics[width=\textwidth]{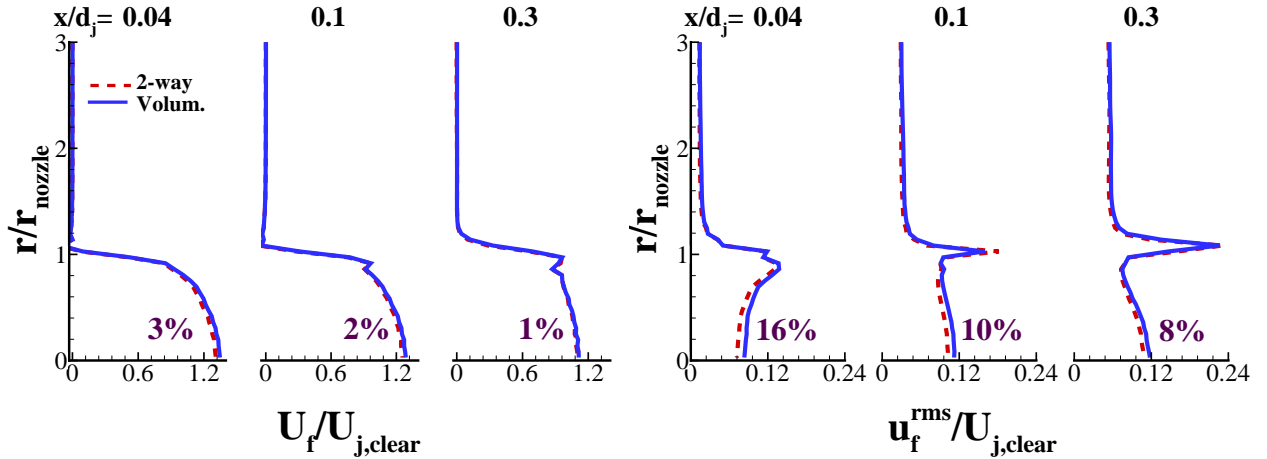}
\caption{Normalized streamwise mean (left) and r.m.s. (right) velocities of the carrier phase for case B based on two different couplings.} \label{fig:f_100}
\end{figure}

\begin{figure}
\setlength{\unitlength}{0.012500in}
\centering
\includegraphics[width=\textwidth]{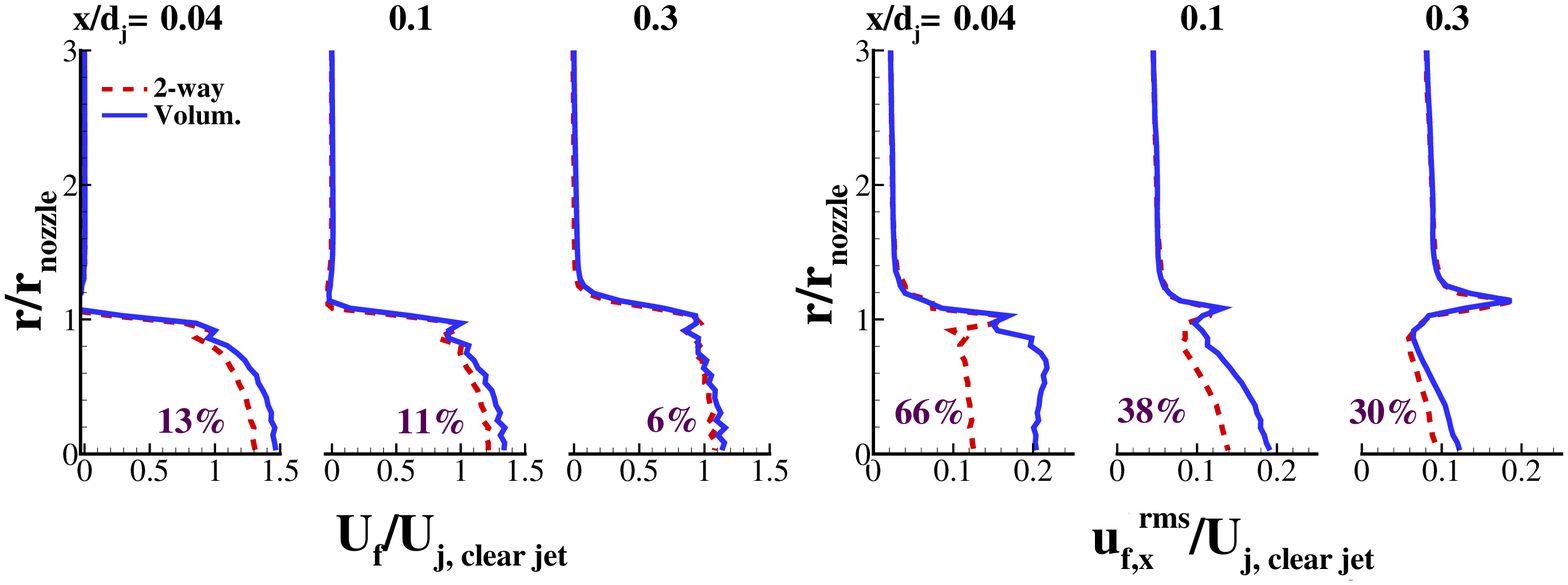}
\caption{Normalized streamwise mean (left) and r.m.s. (right) velocities of the carrier phase for case C based on two different couplings.} \label{fig:f_400}
\end{figure}

\begin{figure}
\setlength{\unitlength}{0.012500in}
\centering
\includegraphics[width=\textwidth]{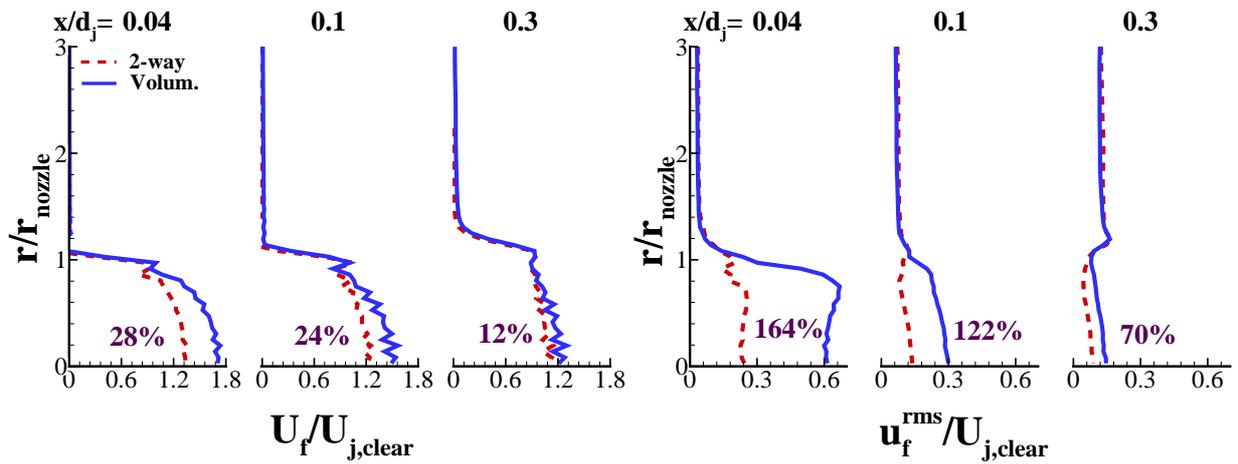}
\caption{Normalized streamwise mean (left) and r.m.s. (right) velocities of the carrier phase for case D based on two different couplings.} \label{fig:f_800}
\end{figure}

The purpose of this section is to study the volumetric displacement effects of dispersed phase onto the characteristics of the carrier phase compared with the predictions of common two-way Euler-Lagrange approaches where this effect is neglected. As the regime in cases B-E is dense, therefore inter-particle collision is also required to account for. Soft-sphere model of \cite{Cundall} is employed here. Collision parameters and formulation can be found in our earlier works, e.g., \cite{Finn2016,Pakseresht2017_ASME}. A comparison between prediction of volumteric and standard two-way couplings on the mean and r.m.s. velocities of the carrier phase corresponding to cases B to D is performed here. Figures \ref{fig:f_100}-\ref{fig:f_800} depict the LES-PP results of cases B-D respectively for mean and r.m.s. velocities of the carrier phase. For case B as shown in Figure \ref{fig:f_100}, slight difference exits between these two approaches with the maximum of 3\% and 16\% increase in the mean and r.m.s. velocities predicted by the volumetric coupling right at the nozzle exit. However, this difference becomes more pronounced and distinguishable with increasing particles' loading such as those in cases C and D plotted in Figures \ref{fig:f_400} and \ref{fig:f_800} respectively. As illustrated in these plots, volumetric coupling predicts higher mean and r.m.s. velocities remarkably compared to the standard two-way coupling.  Furthermore, one can see that the difference between two-way and volumetric couplings decreases with nozzle distance in all cases which could be conjectured due to spread and enterainment of the particulate jet which in turn decreases the local volume fraction of the dispersed phase so does the difference between these two approaches farther away from the nozzle exit.

\begin{figure}
\begin{minipage}{75mm}
\begin{center}
\includegraphics[width=\textwidth]{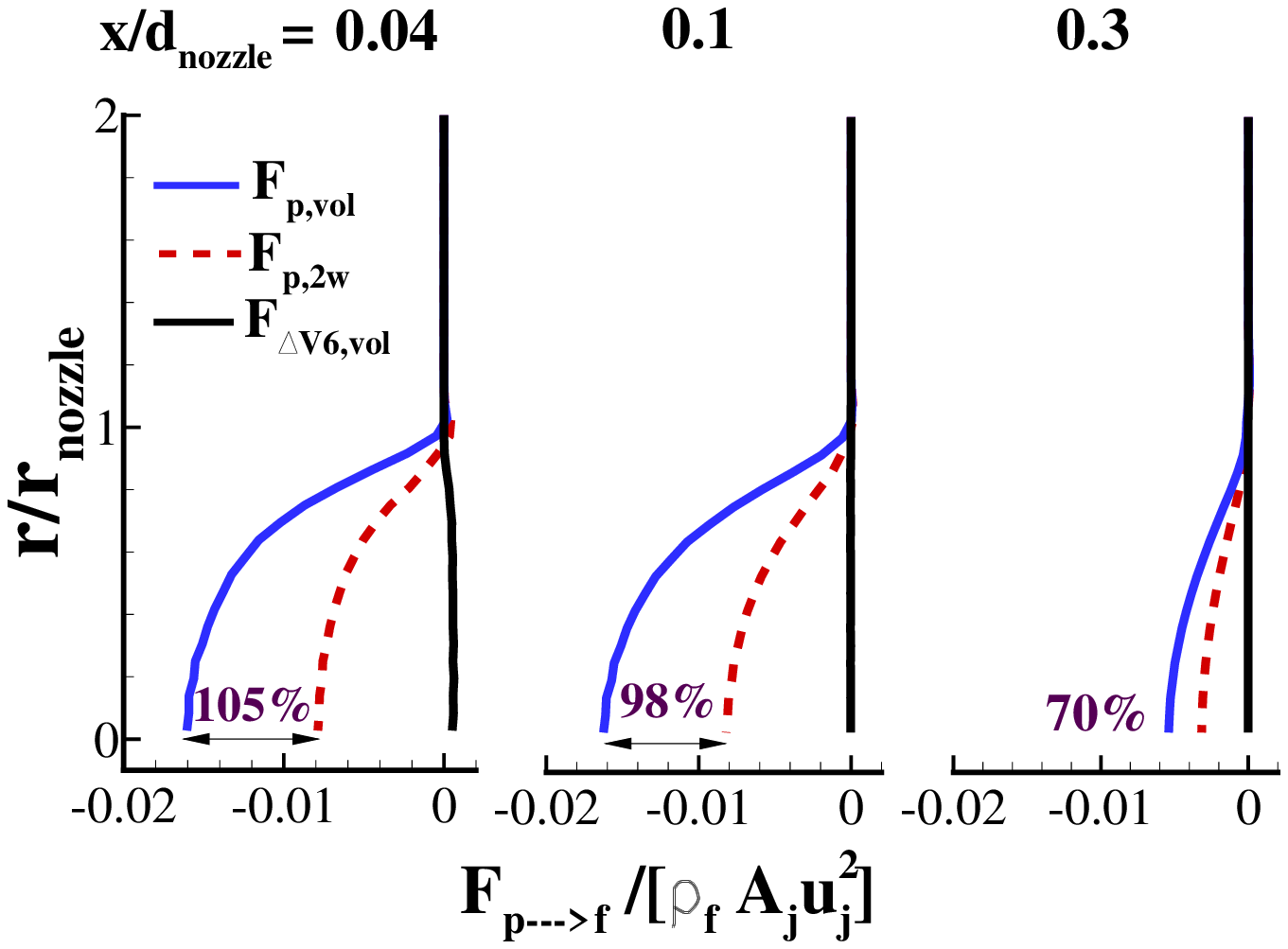}
\caption{Contribution of particle reaction forces by two different coupling formulations normalized by the jet momentum.}
\label{fig:force_fluid_800}
\end{center}
\end{minipage}
\hfill
\begin{minipage}{75mm}
\begin{center}
\includegraphics[width=\textwidth]{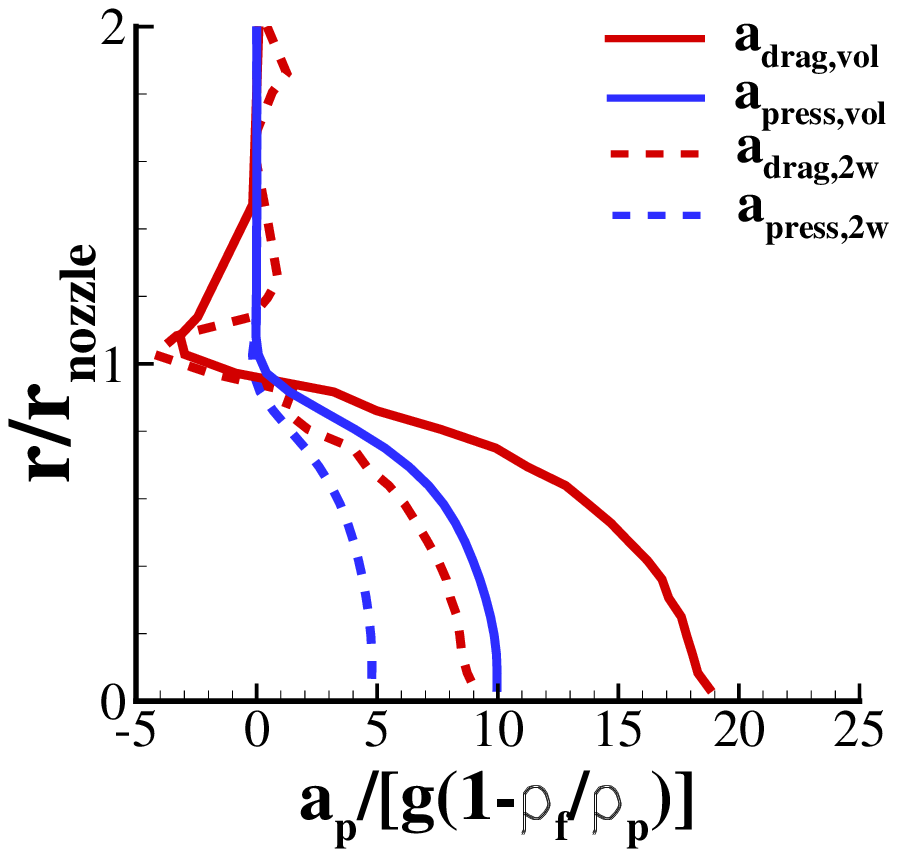}
\caption{Contribution of exerted accelerations onto particles by two different coupling formulations normalized by Buoyancy.}
\label{fig:force_part_800}
\end{center}
\end{minipage}
\end{figure}

\begin{figure}
\setlength{\unitlength}{0.012500in}
\centering
\includegraphics[width=\textwidth]{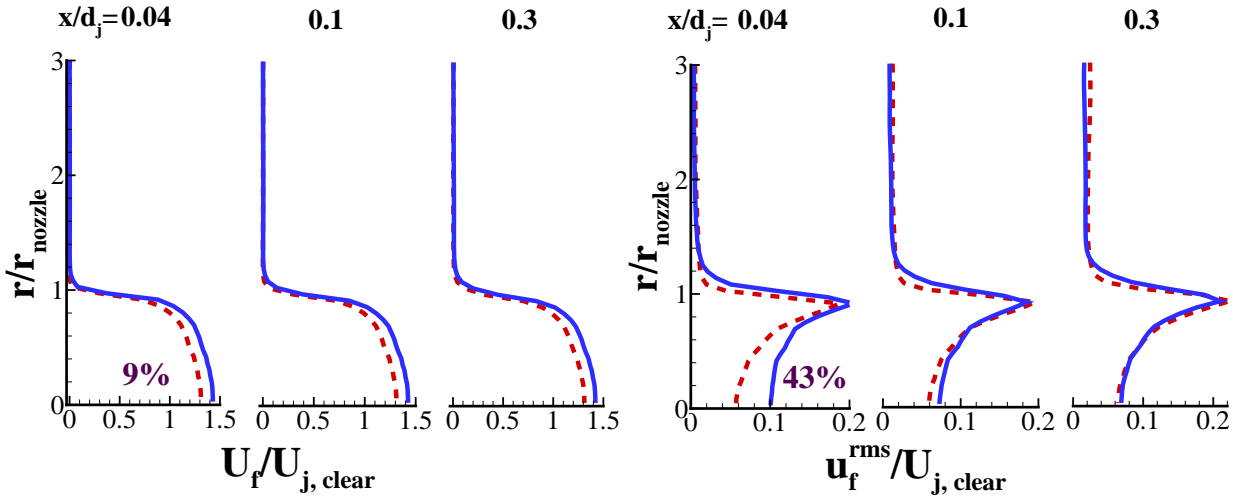}
\caption{Normalized streamwise mean (left) and r.m.s. (right) velocities of the carrier phase for case C based on two different couplings.} \label{fig:u_f_400_equalforces}
\end{figure}

Better insight into the mechanisms of the particle-turbulence interactions involved in the volumetric coupling formulation can be further obtained by doing the force analysis. In the volumetric formulation, it is clear that the carrier phase sees the effects of particles not only through the force closures added to the momentum (standard two-way coupling) but also through the spatio-temporal variations in the carrier phase volume fraction due to presence of particles. It was shown in work of \cite{Cihonski2013} that this is the extra forces arise from volumteric formulation which differentiate this approach than standard two-way coupling. Figure \ref{fig:force_fluid_800} depicts a comparison between two-way and volumetric couplings predictions on the radial profile of particle feedback forces back onto the flow in the near field of the densest jet case (i.e., D). On the same plot, the most dominant force arises in the volumetric formulation observed by \cite{Cihonski2013}, i.e., $\Delta V_6=\rho_f\theta_f \mathbf u_f(\nabla \cdot \mathbf u_f)$ are also plotted. All forces on this plot are normalized by the jet momentum, i.e., $\rho_f A_j U_j^2$ in the stream-wise direction. It is clear that contribution of $\Delta V_6$ is quite insignificant compared to the point-particle force both from volumetric coupling prediction. Furthermore, point particle force predicted by volumetric coupling formulation (i.e., $F_{p,vol}$) is almost twice than that of two-way coupling ($F_{p,2w}$). Likewise, fluid forces exerted onto the particles particularly right at the nozzle shown in Figure \ref{fig:force_part_800} reveal the fact that both drag as well as pressure forces are significantly larger in volumetric coupling prediction than those of standard two-way coupling. This shows that despite the work of \cite{Cihonski2013}, here another mechanism must exist. This can be investigated by looking at the continuity equation used in volumetric coupling formulation where modified density may result in higher fluid velocity compared to the standard two-way coupling. To assess this, one can mask the effect of point-particle forces and purely investigate the volumetric displacement effects by looking at case E where a comparison between one-way coupling and its modification by volumetric formulation is performed and illustrated in Figure \ref{fig:u_f_400_equalforces}. Note that equations for standard one-way coupling can be simply obtained by recalling Equations \ref{mass} and \ref{momentum} and giving $\theta_f=1$ as well as $\mathbf{F}_{p \rightarrow f}=0$. Likewise, modified one-way coupling is achieved by equaling $\mathbf{F}_{p \rightarrow f}=0$ in the same equations, however, keeping spatio-temporal variations in the fluid volume fraction, i.e., $\theta_f\neq1$. As depicted in Figure \ref{fig:u_f_400_equalforces}, a significant difference between these two couplings is observed for both mean and r.m.s. velocities at the nozzle exit. This can be again explained due to the fact that volumetric formulation solves carrier phase equations for less amount of volume (or density) due to presence of particles which in turn increases the corresponding velocity of the carrier phase. This observation is in line with work of \cite{Ferrante2004} where accurate prediction of the drag reduction in the boundary layer over a flat plate was shown to be due to the results of velocity divergence effect, i.e., accounting for spatio-temporal variations in fluid volume fraction. Further analyses particularly the influence of particle's Stokes number can be found in \cite{pakseresht2019_ijmf}.

\section{Summary and Conclusions}
Accurate prediction of a dense spray flow using an Euler-Lagrange approach was presented. To accurately model this flow, volume of the carrier phase displaced by the motion and presence of dispersed phase was taken into account along with the standard two-way coupling point-particle forces employed in typical Euler-Lagrange approaches. To Investigate and quantify the volumetric displacement effects, various volume loadings ranging from dilute up to dense regime (38 \%) were studied. It was observed that for volume loadings of equal and greater than 5\% ($\overline{\theta_p}>0.05$), volumetric coupling predicts higher mean velocity and turbulence intensities for the carrier phase in comparison with the standard two-way coupling. In line with work of \cite{Ferrante2004}, this enhancement was shown to be due to the velocity divergence effect as a result of modified continuity equation in which less amount of density in the region of high void fraction results in higher carrier phase velocity. Finally, the present formulation can be extended to further investigations of other fields such as energy harvesting \citep{kamrani2019,siala2020}, among others, with the aid of micro particles.

\section{Acknowledgements}
The authors acknowledge the Texas Advanced Computing Center (TACC) at The University of Texas at Austin for providing HPC resources that have contributed to the research results reported within this paper. In addition, financial support from the National Aeronautics and Space Administration (NASA) is highly appreciated.

\bibliographystyle{elsarticle-harv}\biboptions{authoryear}
\bibliography{manuscript}

\end{document}